\documentclass[aps,prb,groupedaddress,twocolumn,showpacs,preprintnumbers,
amsmath,amssymb]{revtex4}

\usepackage{graphicx}
\usepackage{dcolumn}
\usepackage{bm}

\begin{document}

\title{Vacancy-induced magnetism in SnO$_{2}$: A density
functional study}

\author{Gul Rahman$^{1}$}
\email[]{grnphysics@yahoo.com}
\author{V\'{i}ctor M. Garc\'{i}a-Su\'{a}rez$^{2}$}
\author{Soon Cheol Hong$^{1}$}

\affiliation{$^{1}$Department of Physics, University of Ulsan,
Ulsan 680-749, Republic of Korea}

\affiliation{$^{2}$Physics Department, Lancaster University,
Lancaster LA1 4YB, United Kingdom}

\date{\today}

\begin{abstract}
We study the magnetic and electronic properties of defects in
SnO$_{2}$ using pseudopotential and all electron methods. Our
calculations show that bulk SnO$_{2}$ is non-magnetic, but it
shows magnetism with a magnetic moment around 4.00 $\mu_{B}$ due
to Sn vacancy (V$_\mathrm{Sn}$). The magnetic moment comes mainly
from O atoms surrounding V$_\mathrm{Sn}$ and Sn atoms, which
couple antiferromagnetically with the O atoms in the presence of
V$_\mathrm{Sn}$. The coupling between different Sn vacancies is
also studied and we find that these defects not only couple
ferromagnetically but also antiferromagnetically and
ferrimagnetically. Our calculations demonstrate that the
experimentally observed giant magnetic moment of transition metal
doped SnO$_{2}$ can be attributed to V$_\mathrm{Sn}$.
\end{abstract}


\maketitle
In many diluted magnetic semiconductor (DMS) the non-magnetic
matrix is a conventional compound semiconductor such as GaAs
\cite{gulref1} or InAs \cite{gulref2}. These DMSs have low
solubility limit and their Curie temperatures ($T_\mathrm{C}$)s
are well below room temperature (RT), which disqualify them for
spintronic devices. In other classes of DMS the transition metal
(TM) is embedded in oxide semiconductors, which are conventionally
known as oxide-DMS (ODMS), such as ZnO with Co or Mn doping
\cite{gulref3,gulref4,gulref5}, TiO$_{2}$ (anatase) with Co
\cite{gulref6} and SnO$_{2}$ with Co \cite{gulref7}. These ODMSs
have large magnetic moments and their $T_\mathrm{C}$s are well
above RT. Therefore these are good candidates for spintronic
devices.

SnO$_{2}$ is a wide band gap material which has been used as a
transparent conducting electrode in solar cells \cite{gulref8} and
flat-panel display \cite{gulref9}. This compound has a rutile
structure with distorted octahedral coordination. Ogale \textit{et
al}. \cite{gulref7} found that Co- doped SnO$_{2}$ not only
exhibits ferromagnetism above RT but also shows a giant magnetic
moment (GMM) of 7.5${\pm}$0.5 $\mu_{B}$/Co, which was reported
to be the first GMM observed in any DMS. The GMM was attributed to
either the cobalt orbital magnetic moment, which was probably not
quenched, or the appearance of some moment on the atoms
surrounding the cobalt in the matrix. Ferromagnetism with high
$T_\mathrm{C}$s was also observed in Fe \cite{gulref10}, Cr
\cite{gulref11}, V \cite{gulref12}, and Ni \cite{gulref13}
doped-SnO$_{2}$. The ferromagnetism in V, Cr, and Ni
doped-SnO$_{2}$ was found to depend on the nature of the substrate
on which the samples were deposited
\cite{gulref11,gulref12,gulref13}. More recently \cite{gulref14},
it has been observed that Mn-doped SnO$_{2}$ also showed very
large magnetic moment i.e. 20.0 $\mu_{B}$ at low doping
concentration which is far above the spin magnetic moment of Mn.

To date, there are few cases where defects are believed to be the
origin of ferromagnetism and each case has a different physical
origin. These cases include CaO \cite{gulref27}, SiC
\cite{gulref28}, CaB$_6$ \cite{gulref29}, and HfO$_{2}$
\cite{gulref30}. Each compound has different crystallographic
environment and local symmetry which gives birth to different
origins of ferromagnetism in these materials. Experimentally it
has been observed these materials show ferromagnetism without
doping of TM, like for example HfO$_{2}$ and CaB$_{6}$. Our case is
quite different than those mentioned above. SnO$_{2}$ has a rutile
structure in which each Sn is surrounded by an oxygen distorted
octahedron. Also, SnO$_{2}$ shows very large magnetic moments
($\sim$ 20 $\mu_{B}$, for Mn-doped SnO$_{2}$) when doped with
TM. However, we will show that it is possible to induce magnetism
in SnO$_{2}$ without doping of TM. Most importantly, the previous
studied materials did not consider the magnetic coupling between
the defects. We study in detail that coupling to prove that
SnO$_{2}$ has magnetism without TM doping.

Considering the defect chemistry of SnO$_{2}$, it has been
concluded that the TM cations are in a +3 ionic state, i.e.
Mn$^{+3}$, Fe$^{+3}$ and Co$^{+3\;}$ \cite{gulref14}. These TMs
have lower valence than Sn for which they substitute and therefore
will influence the defect chemistry of SnO$_{2}$. On the other
hand, in structures where ionic or metallic binding predominates,
as many as half of the cation sites may be vacant or may occupy
interstitial sites \cite{gulref15}. In other words we can say that
vacancies are quite common in many crystals, both in closed-packed
metallic structures as well as in open covalent structures
\cite{gulref16}. Although TM doped-SnO$_{2}$ shows RT
ferromagnetism with large magnetic moments, which is ideal for
spintronic devices, the origin of GMM is still unclear. To find
the physical origin of GMM we considered two kinds of defects i.e.
Sn vacancies (V$_\mathrm{Sn}$) and oxygen vacancies
(V$_\mathrm{O}$). Interestingly we found that V$_\mathrm{Sn}$ has
a magnetic ground state with a large magnetic moment, which
indicates that V$_\mathrm{Sn}$ can be responsible for GMM. On the
other hand V$_\mathrm{O}$ does not show any magnetism. Detailed
energetics and other kinds of defects are outside the scope of the
present work, but they have been studied in detail \cite{gulref31}
before the discovery of ferromagnetism in SnO$_{2}$. So here we
will focus only on V$_\mathrm{Sn}$ and V$_\mathrm{O}$ in
connection with magnetism.

We performed density functional theory calculations using the
SIESTA code \cite{gulref17} with a double-zeta polarized (DZP)
basis set. For the exchange and correlation potential, we used
both the local density approximation (LDA) \cite{gulref18} and the
generalized gradient approximation (GGA) \cite{gulref19}. The
Hamiltonian matrix elements were calculated on a real space grid
defined with a plane-wave energy cutoff of 300 Ry. A
4${\times4\times6}$ Monkhorst-Pack (MP) mesh for one unit cell (6
atoms) and a 3${\times3\times3}$ MP for a 2${\times2\times2}$
supercell (48) atoms were used for \textbf{k} point sampling. The
atomic positions were relaxed until all forces were smaller than
0.05 eV/\AA. To confirm our results, we also carried out
calculations using the full potential linearized augmented plane
wave (FLAPW) \cite{gulref20} with GGA \cite{gulref19}, 48 special
\textbf{k}-points and about 2450 basis functions, which yielded
qualitatively the same results. As a starting point we used our
optimized LDA (GGA) parameters of SnO$_{2}$ which are: $a=4.715 (4.780)$ \AA, $c=3.248 (3.268)$ \AA, and $u=0.307 (0.307)$.
For V$_\mathrm{Sn}$ we considered a defect concentration of
0.0625, which corresponded to a single Sn vacancy in the unit cell
of 48. For V$_\mathrm{O}$ we removed one O atom from the same unit
cell.

\begin{figure}
\includegraphics[width=\columnwidth]{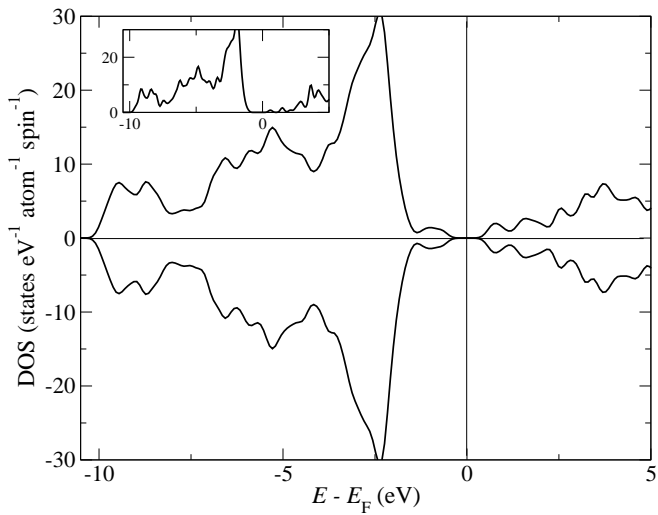}
\caption{\label{fig:epsart} Total DOS for O-defect
(V$_\mathrm{O}$) system calculated with LDA. The inset shows the
total DOS for a pure Sn$_{16}$O$_{32}$ super cell.}
\end{figure}

In a perfect SnO$_{2}$ crystal the Sn is in a 4+ state due to the
donation of its 4 valence electrons to the O$_{2}$ complex, which
is in a 4- state. This results in SnO$_{2}$ as a non-magnetic and
wide band gap material where all bands are occupied. The total DOS
of pure SnO$_{2}$ calculated with and without spin polarization
gives the same result and is shown in the inset of
Fig.~\ref{fig:epsart}. The valence band is formed mainly by O
2\textit{p} orbitals and is full whereas the conduction band is
formed mainly by the Sn 5\textit{s} orbitals and is empty.

On a simple ionic picture, removal of neutral oxygen would lead to
the reduction of tin from Sn$^{4+}$ to Sn$^{2+}$ state, as has
been confirmed by earlier spectroscopic studies \cite{gulref21}.
Dangling bonds cannot be expected since Sn can also exist in a
Sn$^{2+}$ state and it will complete its bonding with the nearby O
atoms, which will result in SnO which is an insulator
\cite{gulref22}. We find that V$_\mathrm{O}$ is a non-magnetic
insulator. The total DOS for V$_\mathrm{O}$ is shown in
Fig.~\ref{fig:epsart}. The oxygen defect creates a defect band
inside the band gap of pure SnO$_{2}$ which does not destroy the
insulating behavior of SnO$_{2}$, but decreases its band gap. This
indicates that the O defect is electronic in nature.

\begin{figure}
\includegraphics[width=\columnwidth]{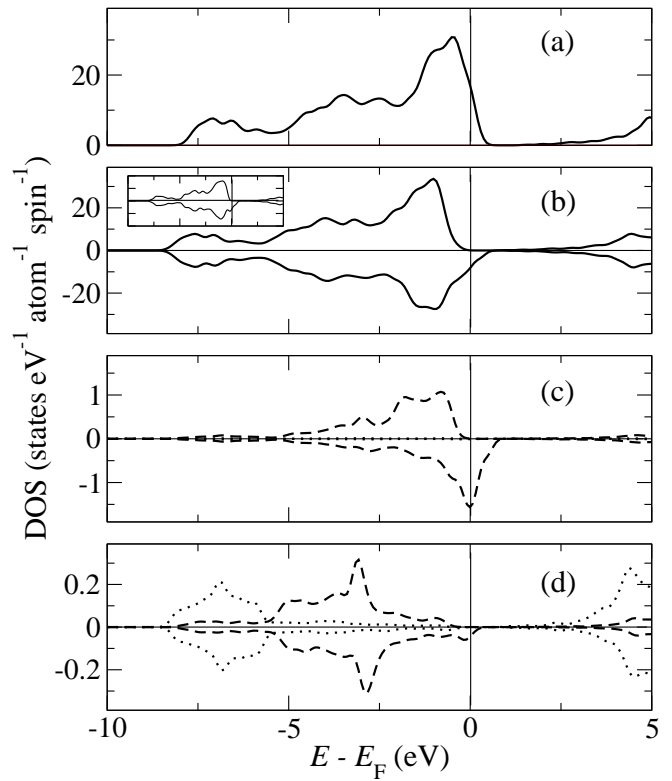}
\caption{\label{fig:epsartb} (a) Total DOS for V$_\mathrm{Sn}$ in
the PM state. (b) Total DOS for V$_\mathrm{Sn}$ in the FM state.
The inset shows the DOS calculated with FLAPW.  Partial DOS of (c)
O and (d) Sn atoms for V$_\mathrm{Sn}$. Dotted and dashed lines
show \textit{s} and \textit{p} orbitals, respectively, whereas the
solid lines show the total DOS.}
\end{figure}

When one Sn vacancy is introduced in the supercell the bond length
of Sn-O is decreased by ${\sim} 0.14$ \AA\/ after relaxation. In
this case holes are created at the O sites. The DOS for
V$_\mathrm{Sn}$ in the paramagnetic case using GGA is shown in
Fig.~\ref{fig:epsartb} (a).

We find that the band structure is identical to pure SnO$_{2}$,
except close to the Fermi energy. Now, a hole band formed by the O
surrounding V$_\mathrm{Sn}$ is created on top of the valence band.
Therefore a cation vacancy removes symmetrized orbitals from the
bonding bands and raises them above the Fermi energy. The DOS
indicates that the electrons are partially localized, compared to
pure SnO$_{2}$ where electrons are completely localized due to
strong bonding, and there is a metallic behavior. The finite
density at the Fermi level indicates that these holes will mediate
a magnetic interaction, if V$_\mathrm{Sn}$ is showing any
magnetism. To see more deeply whether they correspond to spin up
or spin down electrons we spin polarized the band structure. All
our calculations (SIESTA and FLAPW) show that V$_\mathrm{Sn}$ is
more stable in the magnetic (M) state than in the paramagnetic
(PM), with an energy difference, $E_\mathrm{M}-E_\mathrm{PM}$,
which ranges between -0.53 (LDA) and -0.71 eV (GGA).

\begin{figure}
\includegraphics[width=\columnwidth]{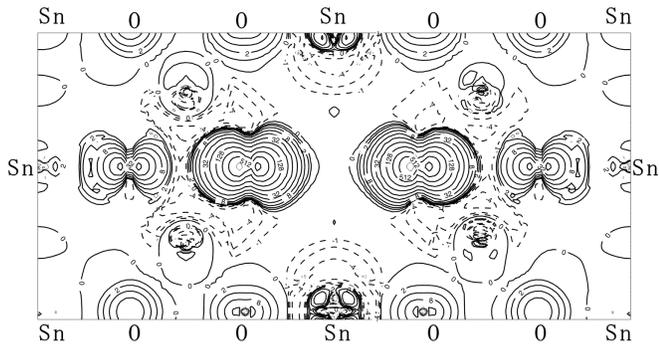}
\caption{\label{fig:epsartc} Spin density contours for
V$_\mathrm{Sn}$ in the (110) plane calculated with GGA using
FLAPW. The lowest contour starts from $2.0\times$ 10$^{-4}$
electrons/(a.u)$^{3}$ and the subsequent lines differ by a factor
of 2.}
\end{figure}

The GGA spin polarized DOS is shown in Fig.~\ref{fig:epsartb} (b).
All the states below -10 eV are spin degenerate and for simplicity
are not shown. However, the situation changes dramatically around
the Fermi level. We see the DOS is spin split, which implies that
the Sn defect induces magnetism in this system. We also find a
large magnetic moment of 3.85 (4.00) $\mu_{B}$ with LDA
(GGA). The Sn defect creates holes at the neighboring O ligands
which are localized for spin up while they are partially localized
for spin down electrons. So when we introduce one Sn defect we are
creating two ligand O holes which in turn give a total magnetic
moment 4.00 $\mu_{B}$, as shown by our ab inito
calculations. We then analyze the PDOS and we find that the O
atoms surrounding the Sn defects are the main ones that contribute
to the magnetism. The PDOS of one of these O atoms is shown in
Fig.~\ref{fig:epsartb} (c). We see the spin down state is
partially occupied, which contributes to the magnetic moment of
V$_\mathrm{Sn}$. For comparison we also show the PDOS of a near Sn
atom (Fig.~\ref{fig:epsartb} (d)), which produces almost no
contribution to the magnetic moment. So the magnetic moment is
mainly attributed to the O atoms due to the dangling bonds that
create unpaired electrons. The total DOS also shows that majority
spins keep the insulating behavior of pure SnO$_{2}$ whereas
minority spins are metallic. Such DOS indicates half metallic
behavior, which is very important for spintronic devices. The LDA
DOS, which is not shown here, produces a similar trend. The FLAPW
also shows that the ground state of V$_\mathrm{Sn}$ is magnetic
with a magnetic moment of 4.00 $\mu_{B}$. For comparison the
total DOS calculated with FLAPW within GGA is shown in the inset
of Fig.~\ref{fig:epsartb} (b). To see the coupling between O and
Sn atoms for the V$_\mathrm{Sn}$ system, we calculated the spatial
projection of the spin densities. The spin contours are shown in
Fig.~\ref{fig:epsartc}.

We see that the Sn atoms are negatively polarized and couple
antiferromagnetically with the O atoms and the spin density
clearly indicates that the magnetic moments are localized at O
sites. These facts are also confirmed by the Mulliken charge
analysis (SIESTA) as well as by the magnetic moment within the MT
spheres (FLAPW). The spin projected densities also show that the
charge of O atoms near V$_\mathrm{Sn}$ spreads towards the defect
due to the dangling bonds and the magnetic moments of such O atoms
are increased compared to those far away from V$_\mathrm{Sn}$. We
can also see that the the magnetic moments of O atoms decrease
rapidly as the distance of O atoms from V$_\mathrm{Sn}$ increases.

\begin{figure}
\includegraphics[width=\columnwidth]{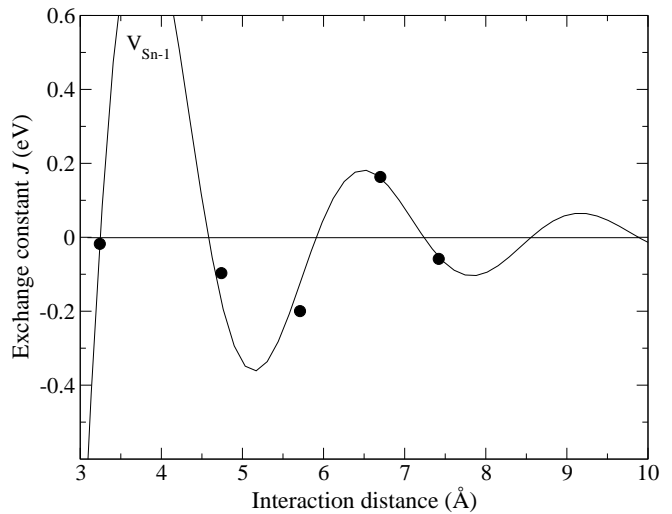}
\caption{\label{fig:epsartd} The exchange coupling constant
$J(eV)$ as a function of V$_\mathrm{Sn}$ separation. The dots are
data points whereas the solid line is the RKKY model fitted with
$k_\mathrm{F}$ = 1.19 \AA\/$^{-1}$}.
\end{figure}

Until now we focused only on single Sn vacancies and we concluded
that V$_\mathrm{Sn}$ is magnetic using FLAPW and SIESTA. However,
to specify wether there is ferromagnetism or antiferromagnetism,
which are cooperative phenomena, there should be at least two
entities in the unit cell that can determine the true magnetic
ground state properties. We therefore move to discuss the coupling
between the V$_\mathrm{Sn}$ defects. We considered two
V$_\mathrm{Sn}$ defects in the supercell and fully relaxed all the
atoms. We varied the distance between the defects and computed the
total energies in the PM, FM and AFM states. The calculated
exchange coupling constants are not exactly equal to the energy
differences $(E_\mathrm{FM}-E_\mathrm{AFM})/\mathrm{vacancy}$
\cite{gulref24} since it is necessary to take into account the
interaction of each vacancy with all periodic images. This is
specially important when the vacancy is at the edges of the unit
cell (e.g. a vacancy located at (1,1,1), in the middle of the
supercell, would be equidistant from 8 vacancies and
$(E_\mathrm{FM}-E_\mathrm{AFM})/\mathrm{vacancy}=8J_{111}$). The
resulting exchange coupling constants were calculated following
the procedure of Zhao \textit{et al}. \cite{gulref24} and the
exchange constants vs defects separation are shown in
Fig.~\ref{fig:epsartd}.

We found that there are two factors that determine the ground
state between any pair of defects, the
V$_\mathrm{Sn}$-V$_\mathrm{Sn}$ separation and the relative
orientation of the anion adjacent to them. There is a strong
competition between FM and AFM direct exchange interactions
depending on the V$_\mathrm{Sn}$-V$_\mathrm{Sn}$ separation when
the direct coupling between the defects is taken into account. The
total magnetic moment of the AFM is zero for all cases except for
V$_\mathrm{Sn-1}$ (as marked in Fig.~\ref{fig:epsartd}) where we
found ferrimagnetic (FIM) coupling. We see the defects are
strongly coupled ferromagnetically when the V$_\mathrm{Sn}$ are
separated by ${\sim}5.5$ \AA\/. Our calculations also show that
most of the total energy differences between FM and AFM are very
large (above RT), which indicate that the high $T_\mathrm{C}$s in
SnO$_{2}$ based DMS come from the Sn vacancies. In Ref.
\cite{gulref23} the authors linked the observed magnetism in
SnO$_{2}$-based systems with the \textit{d$^{0}$} phenomenon. They
extrapolated the magnetic moments vs film thickness and assumed
that if cation vacancies were magnetic then the separation between
them would be 0.5 nm, which is indeed what we observe
\cite{ferromagnetic}.This discloses the origin of GMM and RT
ferromagnetism is Sn vacancies.

Now we address the question why the distance between FM defects is
strongly localized around 5.5 \AA. The Sn vacancy induces a
magnetization which is produced mainly by the O atoms. On the
other hand the Sn atoms couple antiferromagnetically. This
situation resembles a conventional superexchange mechanism, where
the TM has some magnetic moment and the magnetism is mediated by
the intervening O atoms. Wee see that when the angle between O
atoms (O-Sn-O) is 90$^{\circ}$ the separation between the
V$_\mathrm{Sn}$ is strongly localized around 5.5 \AA and the
most stable magnetic configuration is FM. If the angle between O
atoms is increased to 180$^{\circ}$ then the calculations also
show FM behavior. This can not be explained by the typical
Goodenough-Kanamori-Anderson (GKA) because the magnetic vacancies
are coupled by more than one atom. Indeed, according to the GKA
one would expect the AFM behavior to be more stable for
180$^{\circ}$, which is not what we find. The oscillatory behavior
observed in Fig.~\ref{fig:epsartd} is typical of a RKKY
interaction, which appears when the density of defects is high and
the donor states merge with the bottom of the conduction band. By
using the typical RKKY expression \cite{gulref24},

\begin{equation}
J(r)\propto\frac{\sin (2k_\mathrm{F}r)-2k_\mathrm{F}r\cos
(2k_\mathrm{F}r)}{r^4}
\end{equation}

\noindent we fit our data to the RKKY model, which gives a Fermi
wave vector $k_\mathrm{F}\sim 1.19$ \AA $^{-1}$ and explains why
the most stable FM configuration relative to the AFM configuration
corresponds to a distance of 5.5 \AA. The oscillatory behavior
is well defined but the amplitude of the fit is not well
characterized due to the fact that many points are close to the
nodes of the curve. Such oscillatory behavior
was also shown for Co-doped SnO$_{2}$ using first-principles
calculations\cite{gulref32}. A note of caution should be added
however because the high density of states at the Fermi level,
which prompts to localized holes at the vacancies, is not exactly
compatible with the host-like-holes limit where the RKKY model is
supposed to be valid \cite{gulref26}.

\begin{figure}
\includegraphics[width=\columnwidth]{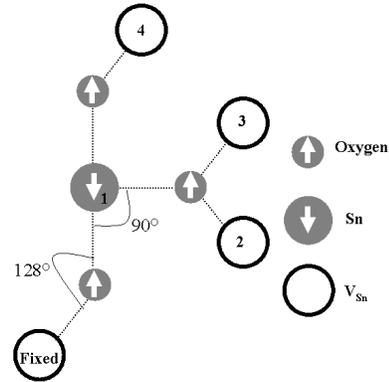}
\caption{\label{fig:epsarte} Mechanism behind the FM and
localization of the two defects in the V$_\mathrm{Sn}$ system. The
arrows are magnetic moments.}
\end{figure}

To further explain why the defects couple ferromagnetically,
antiferromagnetically or ferrimagnetically, we take into account
the angle between them. They may be connected directly (as shown
above) or through a different path followed by different angles
between them as shown in Fig.~\ref{fig:epsarte}.

In Fig.~\ref{fig:epsarte} the V$_\mathrm{Sn}$ marked "Fixed"
remained fixed while 1, 2, 3, and 4 are Sn vacancies at different
sites. We see all the defects are connected through an O-Sn-O
path, i.e. V$_\mathrm{Fixed}$-O-Sn-O-V$_{2-4}$, except one case
where defects share a common oxygen atom. When we consider the
coupling between V$_\mathrm{Fixed}$ and V$_{2}$
(V$_\mathrm{Fixed}$-O-Sn-O-V$_{2}$), for example, all the Sn atoms
at site 1, 3, and 4 are negatively polarized and the ground state
may be FM or AFM. If the defects share a common oxygen, i.e.
V$_\mathrm{Fixed}$-O-V$_{1}$, then the ground state may be PM or
FIM, depending on the distance between the defects. When the
distance is small the ground state is PM. On the other hand, when
we increase the distance or the angle between the defects, the
ground state is found to be FIM, as marked (V$_\mathrm{Sn-1}$) in
Fig.~\ref{fig:epsartd}. The charge transfer is larger for the
shared oxygen atom compared with other oxygens and such atom gives
a very large magnetic moment (${\sim}$ 2.00 $\mu_{B}$) in the FM
calculation. The AFM calculation converges to FIM and gives 2.00
$\mu$$_{B}$/unit cell (1.00 $\mu_{B}$/defect). The structural
distortion was found to be larger for this case as compared to
other cases. The angle between V$_\mathrm{Fixed}$-O-V$_{1}$ is
128$^{\circ}$, which is a typical angle for which FIM has already
been observed in spinals \cite{gulref25}. More recently it has
been found theoretically that Fe-doped SnO$_{2}$ also showed FIM
(1.00 $\mu_{B}$ /Fe) when the Fe impurities connect through O
atoms \cite{gulref26}. On the other hand, if we decrease the
separation between defects while sharing a common O atom then the
total energy difference $E_\mathrm{FM}-E_\mathrm{AFM}$ becomes
very small and PM becomes the most stable state.

In conclusion, we used density functional theory within the local
density approximation (LDA) and the generalized gradient
approximation (GGA) to understand the origin of giant magnetic
moments (GMM) in SnO$_{2}$. We observed that O vacancy
(V$_\mathrm{O}$) is non-magnetic, but Sn vacancy (V$_\mathrm{Sn}$)
showed ferromagnetic behavior with a large magnetic moment. Both
all electron and pseudopotential approaches gave the same results.
The calculated magnetic moment due to V$_\mathrm{Sn}$ in SnO$_{2}$
was mainly contributed by the O atoms surrounding V$_\mathrm{Sn}$
whereas the Sn atoms coupled antiferromagnetically. We also
calculated the magnetic coupling between Sn vacancies and we
showed that these defects can couple ferromagnetically,
antiferromagnetically and ferrimagnetically. Strong ferromagnetic
coupling between Sn defects was found when they were separated by
${\sim}$ 5.5 \AA. We also showed that the exchange coupling
between these defects oscillates and such oscillatory behavior was
attributed to the typical RKKY oscillation. The results of the
extensive LDA calculations are consistent with RT ferromagnetism
and the observed GMM, provided that an extremely high
concentration of Sn vacancies of 12\% (i.e. 2 vacancies per 16 Sn
sites) would be present.

This work was supported by Korea Science and Engineering
Foundation and by the European Commission. We would like to thank Jaime Ferrer and
Stefano Sanvito for useful discussions.

\bibliography{gul}

\begin{thebibliography}{33}
\bibitem[{fer()}]{ferromagnetic}
\bibinfo{note}{The most stable ferromagnetic configuration corresponds to the
  first point, where the separation between vacancies is smallest, but since
  this case is very close in energy to the antiferromagnetic case, it is
  expected to be much less important than the higly stable FM point.}
\end{thebibliography}

\end{document}